\pdfoutput=1

\documentclass[twocolumn,aps,pra,superscriptaddress]{revtex4}

\usepackage{graphicx}
\usepackage{epsfig}
\usepackage{amsfonts}
\usepackage{amsmath}
\usepackage{amssymb}
\usepackage{color}
\usepackage{multirow}
\usepackage[colorlinks=true,citecolor=blue,urlcolor=blue]{hyperref}

\newcommand{\ket}[1]{\vert #1 \rangle}

\def\endproof{\vrule height6pt width6pt depth0pt} 

\begin{document}

%%%%%%%%%%%%%%%%%%%%%%%%%%%%%%%%%%%%%%%%%%%%%%%%%%%%%%%%%%%%%%%%%%%

\title{Bounding quantum theory with the exclusivity principle in a two-city experiment}

%%%%%%%%%%%%%%%%%%%%%%%%%%%%%%%%%%%%%%%%%%%%%%%%%%%%%%%%%%%%%%%%%%%

%dambrosiovincenzo@gmail.com,fabio.sciarrino@uniroma1.it,boure@physto.se,fizmn@univ.gda.pl
%Nawareg:
%fizmn@univ.gda.pl

%%%%%%%%%%%%%%%%%%%%%%%%%%%%%%%%%%%%%%%%%%%%%%%%%%%%%%%%%%%%%%%%%%%

\author{Mohamed Nawareg}
 \affiliation{Department of Physics, Stockholm University, S-10691
 Stockholm, Sweden}
 \affiliation{Instytut Fizyki Teoretycznej i Astrofizyki
Uniwersytet Gda\'nski, PL-80-952 Gda\'nsk, Poland}

\author{Fabrizio Bisesto}
 \affiliation{Dipartimento di Fisica, ``Sapienza''
 Universit\`{a} di Roma, I-00185 Roma, Italy}

\author{Vincenzo D'Ambrosio}
 %\email{dambrosiovincenzo@gmail.com}
 \affiliation{Dipartimento di Fisica, ``Sapienza''
 Universit\`{a} di Roma, I-00185 Roma, Italy}

\author{Elias Amselem}
 \affiliation{Department of Physics, Stockholm University, S-10691
 Stockholm, Sweden}

\author{Fabio~Sciarrino}
 %\email{fabio.sciarrino@uniroma1.it}
 %\homepage{http://quantumoptics.phys.uniroma1.it}
 \affiliation{Dipartimento di Fisica, ``Sapienza''
 Universit\`{a} di Roma, I-00185 Roma, Italy}
 \affiliation{Istituto Nazionale di Ottica (INO-CNR),
 Largo E. Fermi 6, I-50125 Firenze, Italy}

\author{Mohamed~Bourennane}
 %\email{boure@physto.se}
 \affiliation{Department of Physics, Stockholm University, S-10691
 Stockholm, Sweden}

\author{Ad\'an~Cabello}
 %\email{adan@us.es}
 \affiliation{Departamento de F\'{\i}sica Aplicada II, Universidad de
 Sevilla, E-41012 Sevilla, Spain}

%%%%%%%%%%%%%%%%%%%%%%%%%%%%%%%%%%%%%%%%%%%%%%%%%%%%%%%%%%%%%%%%%%%

\date{\today}
%First version: April 28, 2013 (Sevilla).
%This Version: October 13, 2013 (Sevilla).

%%%%%%%%%%%%%%%%%%%%%%%%%%%%%%%%%%%%%%%%%%%%%%%%%%%%%%%%%%%%%%%%%%%

\maketitle

%%%%%%%%%%%%%%%%%%%%%%%%%%%%%%%%%%%%%%%%%%%%%%%%%%%%%%%%%%%%%%%%%%%

{\bf Why do correlations between the results of measurements performed on physical systems violate Bell \cite{Bell64,CHSH69,FC72,ADR82,WJSWZ98,RMSIMW01,GMRWKBLCGNUZ13} and other non-contextuality inequalities \cite{KCBS08,Cabello08,Kirchmair09,Amselem09,Lapkiewicz11,Amselem12,Zhang13,D'Ambrosio13,AACB13} {\em up to some specific limits but not beyond them}?
The answer may follow from the observation that in quantum theory, unlike in other theories, whenever there is an experiment to measure $A$ simultaneously with $B$, another to measure $B$ with $C$, and another to measure $A$ with $C$, there is always an experiment to measure all of them simultaneously \cite{Specker60,LSW11}. This property implies that quantum theory satisfies a seemingly irrelevant restriction called the exclusivity (E) principle: the sum of the probabilities of any set of {\em pairwise} exclusive events cannot exceed~1 \cite{Cabello13,CDLP13,Yan13,Cabello13b,ATC13}, which, surprisingly, explains the set of quantum correlations in some fundamental scenarios \cite{Cabello13,ATC13}. A problem opened in \cite{Cabello13} is whether the E principle explains the maximum quantum violation of the Bell-CHSH inequality \cite{Bell64,CHSH69} and quantum correlations in other scenarios. Here we show experimentally that the E principle imposes an upper bound to the violation of the Bell-CHSH inequality that matches the maximum predicted by quantum theory. For that, we use the result of an {\em independent} experiment testing a specific non-contextuality inequality \cite{Yan13,Cabello13b,ATC13}. We perform both experiments: the Bell-CHSH inequality experiment on polarization-entangled states of pairs of photons in a laboratory in Stockholm and, to demonstrate independence, the non-contextuality inequality experiment on single photons' orbital angular momentum states in a laboratory in Rome. The observed results provide the first experimental evidence that the E principle determines the limits of quantum correlations for both scenarios and prove that hypothetical super-quantum violations for either experiment would violate the E principle. This supports the conclusion that the E principle captures a fundamental limitation of nature. If this is true, much of quantum theory trivially follow from merely taking the E principle to be a fundamental truth, and various information-theoretic postulates are also simplified and/or strengthened.}

%%%%%%%%%%%%%%%%%%%%%%%%%%%%%%%%%%%%%%%%%%%%%%%%%%%%%%%%%%%%%%%%%%%

%\pacs{03.65.Ta, 03.65.Ud, 42.50.Xa, 42.50.Tx}
%03.65.Ta: Foundations of quantum mechanics; measurement theory
%03.65.Ud: Entanglement and quantum nonlocality
%(e.g. EPR paradox, Bell's inequalities, GHZ states, etc.)
%42.50.Xa: Optical tests of quantum theory
%42.50.Tx Optical angular momentum and its quantum aspects

%%%%%%%%%%%%%%%%%%%%%%%%%%%%%%%%%%%%%%%%%%%%%%%%%%%%%%%%%%%%%%%%%%%

Quantum theory (QT) is the most successful scientific theory of all times. However, the reason for its success is not clearly understood, as it is not known how to derive QT from fundamental physical principles. Recently, this problem has been addressed using different approaches, including reconstructing QT from information-theoretic axioms \cite{Hardy01,Hardy11,DB11,MM11,CDP11,MMAP13} and looking for principles for quantum non-local correlations \cite{V05,PPKSWZ09,NW09,FSABCLA13}. One of these approaches \cite{Cabello11,Cabello13,CDLP13,Yan13,Cabello13b,ATC13} seeks principles for explaining the specific way in which QT is contextual, i.e., the manner it violates Bell and other non-contextuality (NC) inequalities.

For years, violations of NC inequalities have been used to emphasize the conflict between QT and local hidden variable theories \cite{Bell64} and between QT and non-contextual hidden variable theories \cite{KCBS08,Cabello08} (i.e., theories in which measurement outcomes are determined before measurements are performed and are independent of which combination of jointly measurable observables is considered). The observation that the {\em specific way} in which QT violates NC inequalities may contain valuable information about the principles of QT is relatively recent \cite{Cabello11}.

The explanation of the limits of the quantum violations of the different NC inequalities may follow from an observation made long ago \cite{Specker60}: In QT, whenever there is one experiment that jointly measures observables $A$ and $B$, one experiment that jointly measures $B$ and $C$, and one experiment that jointly measures $A$ and $C$, there is always one experiment that jointly measures $A$, $B$ and $C$. This observation implies (see the proof in the Supplementary Material) a condition that can be taken as a principle, the exclusivity (E) principle: {\em the sum of the probabilities of any set of pairwise exclusive events cannot exceed~1}. By `event' we mean the outcome of some experimental test, such that two events are mutually exclusive when they are represented by different outcomes of the same test.

So far, the E principle has shown an extraordinary predictive power: For certain experimental scenarios, the E principle singles out the entire set of quantum correlations \cite{Cabello13,ATC13}. Furthermore, for all other scenarios it singles out the entire set of quantum correlations under the assumption that the correlations for a specific independent experiment are given by QT \cite{Yan13,ATC13}. Here we capitalize on the recent theoretical developments established in \cite{ATC13} to experimentally explain the maximum violation of the Bell-CHSH inequality. We will show that the quantum maximum of the Bell-CHSH inequality is determined by the E principle given the result of a specific independent experiment. Moreover, the experimental observation of the maximum value predicted by QT for either of the two experiments directly eliminates the possibility of super-quantum correlation in the other, by virtue of the E principle.

%%%%%%%%%%%%%%%%%%%%%%%%%%%%%%%%%%%%%%%%%%%%%%%%%%%%%%%%%%%%%%%%%%%
% Fig. 1
%%%%%%%%%%%%%%%%%%%%%%%%%%%%%%%%%%%%%%%%%%%%%%%%%%%%%%%%%%%%%%%%%%%

\begin{figure}[bt]
\centering
\includegraphics[scale=0.61]{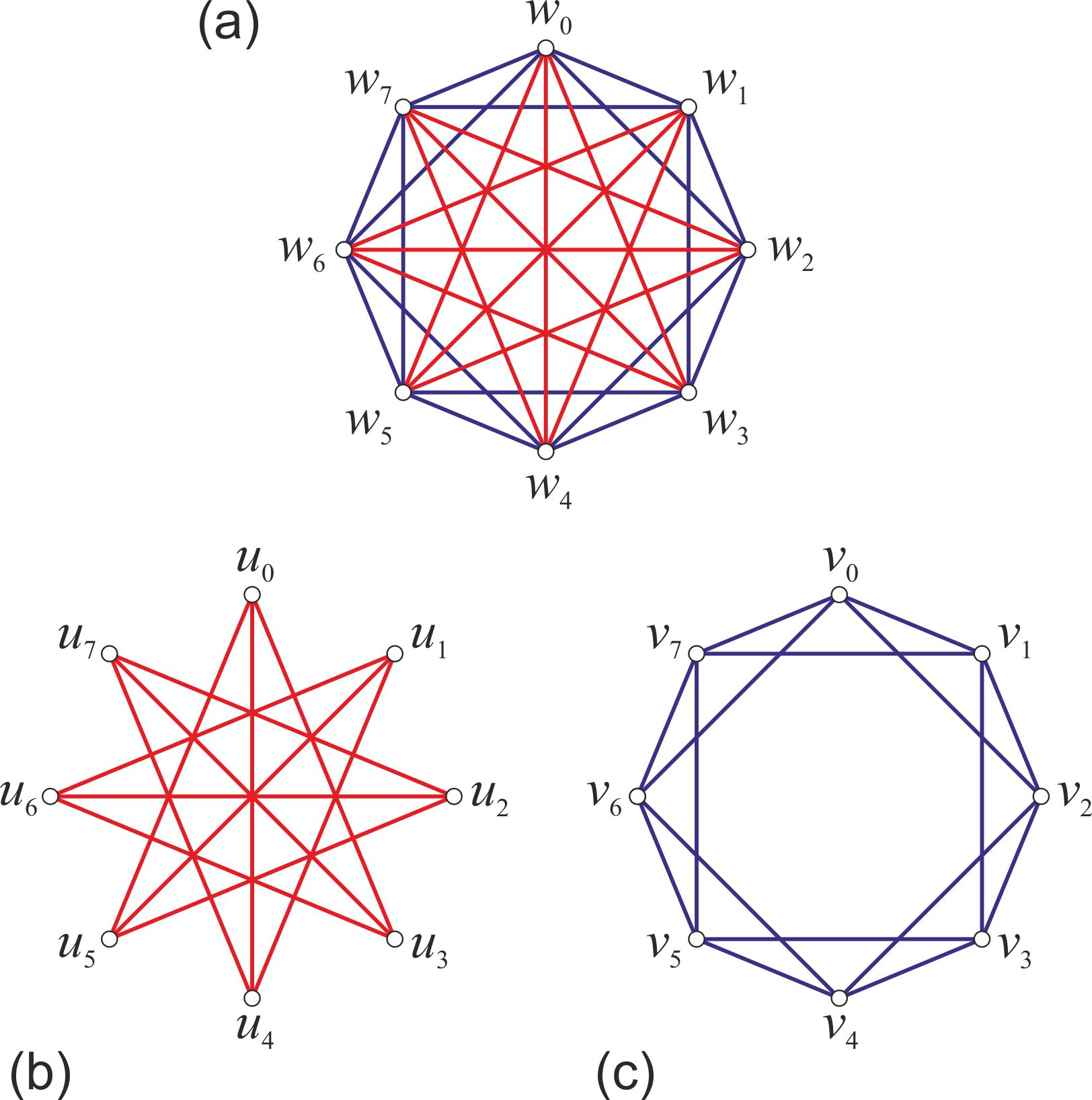}
\caption{\label{Fig1}(a) Graph representing the mutual exclusivity relations between 8 pairwise exclusive events $w_i$. (b) Graph representing the mutual exclusivity relations between the 8 events in $S$. (c) Graph representing the mutual exclusivity relations between the 8 events in $R$. Important for the argumentation is that (b) and (c) are vertex-transitive and mutually complementary graphs.}
\end{figure}

%%%%%%%%%%%%%%%%%%%%%%%%%%%%%%%%%%%%%%%%%%%%%%%%%%%%%%%%%%%%%%%%%%%

The idea can be explained as follows: Consider, for example, 8 pairwise mutually exclusive events $w_i$, with $i=0,\ldots,7$.
%(a trivial example is: $w_0=0,0,0|A,B,C$ denoting ``coins $A$, $B$ and $C$ are tossed and outcomes are: head, head and head, respectively'', $w_1=0,0,1|A,B,C$,\ldots, $w_7=1,1,1|A,B,C$).
The E principle states that the sum of their probabilities satisfy $\sum_{i=0}^7 p(w_i) \stackrel{\mbox{\tiny{E}}}{\leq} 1$. Each event can be represented as a node in the graph of Fig.~\ref{Fig1}(a) in which edges represent mutual exclusivity.

Now imagine two {\em independent} experiments ${\cal S}$ and ${\cal R}$ and suppose that event $w_i$ is defined as the one in which event $u_i$ occurs in experiment ${\cal S}$ and event $v_i$ occurs in experiment ${\cal R}$. Independence implies that
\begin{equation}
 p(w_i) = p(u_i) p(v_i).
\end{equation}
Therefore, the E principle establishes that
\begin{equation}
 W_1 \equiv \sum_{i=0}^7 p(u_i) p(v_i) \stackrel{\mbox{\tiny{E}}}{\leq} 1.
 \label{W1}
\end{equation}

Now suppose that experiments ${\cal S}$ and ${\cal R}$ are devised in such a way that the graph of Fig.~\ref{Fig1}(b) represents the relations of mutual exclusivity between the events $u_i$ and the graph of Fig.~\ref{Fig1}(c) represents the relations of mutual exclusivity between the events $v_i$. Note that every $w_i$ and $w_j$ are mutually exclusive because either $u_i$ and $u_j$ are mutually exclusive, or else $v_i$ and $v_j$ are mutually exclusive.

Let's define
\begin{subequations}
\begin{align}
 &S \equiv \sum_{i=0}^7 p(u_i), \\
 &R \equiv \sum_{i=0}^7 p(v_i).
\end{align}
\end{subequations}
%Notice that the fact that each of the graphs in Fig.~\ref{Fig1}(b) and (c) is vertex-transitive and that the graph in %Fig.~\ref{Fig1}(c) is the complement of the graph of Fig.~\ref{Fig1}(b) allows us to play the following trick:
Since ${\cal S}$ and ${\cal R}$ are independent, if we rotate Fig.~\ref{Fig1}(b) by $k \pi/4$ rad clockwise, with $k=0,\ldots,7$, and make the specular reflection with respect to the axis $v_0$-$v_4$ of Fig.~\ref{Fig1}(c) $m$ times, with $m=0,1$, and then we merge the two resulting figures, we end up with the graph in Fig.~\ref{Fig1}(a) but now representing the mutual exclusivity relations between the events $(u_i,v_j)$. This gives us 16 conditions like (\ref{W1}) (which corresponds to $k=0$, $m=0$). For instance, for $k=1$, $m=0$, we have
\begin{equation}
 W_2 \equiv \sum_{i=0}^7 p(u_i) p(v_{i+1}) \stackrel{\mbox{\tiny{E}}}{\leq} 1,
 \label{W2}
\end{equation}
where the sum in $i+1$ is taken modulo 8.
Summing all these 16 inequalities and dividing by 2, we obtain
\begin{equation}
 S \times R \stackrel{\mbox{\tiny{E}}}{\leq} 8.
 \label{result}
\end{equation}

%%%%%%%%%%%%%%%%%%%%%%%%%%%%%%%%%%%%%%%%%%%%%%%%%%%%%%%%%%%%%%%%%%%
% Table 1
%%%%%%%%%%%%%%%%%%%%%%%%%%%%%%%%%%%%%%%%%%%%%%%%%%%%%%%%%%%%%%%%%%%

\begin{table}[t]
\centering
\begin{tabular}{c c c c}
\hline \hline
Event & $p_{|\psi \rangle}(a,b\,|\,i,j)$ & Experimental value & Expected \\ [0.5ex]
\hline
$u_0$ & $p_{|\psi \rangle}(1,1\,|\,0,0)$ &   $0.4262 \pm 0.0031$ & $0.4267$ \\
$u_1$ & $p_{|\psi \rangle}(1,1\,|\,1,0)$ &   $0.4239 \pm 0.0057$ & $0.4267$ \\
$u_2$ & $p_{|\psi \rangle}(1,-1\,|\,1,1)$ &  $0.4313 \pm 0.0069$ & $0.4267$ \\
$u_3$ & $p_{|\psi \rangle}(-1,-1\,|\,0,1)$ & $0.4319 \pm 0.0058$ & $0.4267$ \\
$u_4$ & $p_{|\psi \rangle}(-1,-1\,|\,0,0)$ & $0.4259 \pm 0.0031$ & $0.4267$ \\
$u_5$ & $p_{|\psi \rangle}(-1,-1\,|\,1,0)$ & $0.4257 \pm 0.0045$ & $0.4267$ \\
$u_6$ & $p_{|\psi \rangle}(-1,1\,|\,1,1)$ &  $0.4226 \pm 0.0028$ & $0.4267$ \\
$u_7$ & $p_{|\psi \rangle}(1,1\,|\,0,1)$ &   $0.4260 \pm 0.0031$ & $0.4267$ \\ \hline
& $S$ & $3.413 \pm 0.013$ & $3.4142$ \\
\hline \hline
\end{tabular}
\caption{\label{Table1}Notation and experimental results for the Bell-CHSH inequality experiment corresponding to the graph of Fig.~\ref{Fig1}(b). $p_{|\psi \rangle}(a, b \,|\, i, j)$ denotes the joint probability of the event ``outcome $a$ is obtained when test $i$ is performed on the first particle and outcome $b$ is obtained when test $j$ is performed on the second particle'' when the initial state is $|\psi \rangle$.}
\end{table}

%%%%%%%%%%%%%%%%%%%%%%%%%%%%%%%%%%%%%%%%%%%%%%%%%%%%%%%%%%%%%%%%%%%

Now notice that the Bell-CHSH inequality can be written as
\begin{equation}
 S \equiv \sum_{i=0}^7 p(u_i) \stackrel{\mbox{\tiny{ NCHV}}}{\leq} 3,
\label{London}
\end{equation}
where events $u_i$ are defined in table~\ref{Table1} and have the relations of mutual exclusivity of Fig.~\ref{Fig1}(b).
$\stackrel{\mbox{\tiny{ NCHV}}}{\leq} 3$ indicates that the maximum value of $S$ for non-contextual hidden variable theories is~$3$. In QT, the maximum value of $S$ is $2+\sqrt{2} \approx 3.4142$.

Result (\ref{result}) indicates that, if we perform an experiment to measure $R$ and take its result, $R_{\rm exp}$, as a lower bound for $R$, the E principle leads to the following upper bound for $S$:
\begin{equation}
 S \stackrel{\mbox{\tiny{E}}}{\leq} 8/R_{\rm exp}.
 \label{Sbound}
\end{equation}

To experimentally measure $R$ we need 8 events with the relations of mutual exclusivity represented in Fig.~\ref{Fig1}(c).
A set of events $v_i$ with this property is given in table~\ref{Table2}. These events satisfy the following NC inequality:
\begin{equation}
 R=\sum_{i=0}^7 p (v_i) \stackrel{\mbox{\tiny{ NCHV}}}{\leq} 2,
 \label{Paris}
\end{equation}
where the upper bound follows from the fact that 2 is the maximum number of events that can have probability~1 while satisfying the relations of exclusivity in Fig.~\ref{Fig1}(c). In QT, the maximum value of $R$ is $8-4 \sqrt{2} \approx 2.3431$ and requires a quantum system of dimension 5 or higher \cite{Cabello13b}.

Result (\ref{result}) shows that, if we perform an experiment to measure $S$ and take its result, $S_{\rm exp}$, as a lower bound for $S$, the E principle leads to the following upper bound for $R$:
\begin{equation}
 R \stackrel{\mbox{\tiny{E}}}{\leq} 8/S_{\rm exp}.
 \label{Rbound}
\end{equation}

The aim of this work is to exploit relation (\ref{Sbound}) to experimentally obtain the tightest possible upper bound for $S$ and exploit relation (\ref{Rbound}) to experimentally obtain the tightest possible upper bound for $R$.

%%%%%%%%%%%%%%%%%%%%%%%%%%%%%%%%%%%%%%%%%%%%%%%%%%%%%%%%%%%%%%%%%%%
% Table 2
%%%%%%%%%%%%%%%%%%%%%%%%%%%%%%%%%%%%%%%%%%%%%%%%%%%%%%%%%%%%%%%%%%%

\begin{table}[htbp]
\centering
\begin{tabular}{cccc}
\hline \hline
Event & $p_{|\phi\rangle}(0,0,1|i-2,i-1,i)$ & Experimental value & Expected \\ [0.5ex]
\hline
$v_0$ & $p_{|\phi\rangle}(0,0,1|6,7,0)$ & $0.2809 \pm 0.0038$ & $0.2929$ \\
$v_1$ & $p_{|\phi\rangle}(0,0,1|7,0,1)$ & $0.2854 \pm 0.0038$ & $0.2929$ \\
$v_2$ & $p_{|\phi\rangle}(0,0,1|0,1,2)$ & $0.2857 \pm 0.0038$ & $0.2929$ \\
$v_3$ & $p_{|\phi\rangle}(0,0,1|1,2,3)$ & $0.3110 \pm 0.0039$ & $0.2929$ \\
$v_4$ & $p_{|\phi\rangle}(0,0,1|2,3,4)$ & $0.2983 \pm 0.0038$ & $0.2929$ \\
$v_5$ & $p_{|\phi\rangle}(0,0,1|3,4,5)$ & $0.2833 \pm 0.0036$ & $0.2929$ \\
$v_6$ & $p_{|\phi\rangle}(0,0,1|4,5,6)$ & $0.2810 \pm 0.0036$ & $0.2929$ \\
$v_7$ & $p_{|\phi\rangle}(0,0,1|5,6,7)$ & $0.3095 \pm 0.0038$ & $0.2929$ \\
\hline
& $R$ & $2.335 \pm 0.011$ & $2.3431$ \\ \hline \hline
\end{tabular}
\caption{\label{Table2}Notation and experimental results for the NC inequality experiment corresponding to the graph of Fig.~\ref{Fig1}(c). $p_{|\phi\rangle}(0,0,1\,|\,i-2,i-1,i)$ denotes the probability of the event ``outcomes $0,0,1$ are obtained when the jointly measurable observables $i-2,i-1,i$ are measured'' when the initial state is $|\phi \rangle$. In the experiment we have first checked that events $1\,|\,i-2$, $1\,|\,i-1$ and $1\,|\,i$ are pairwise mutually exclusive and then used this to conclude that $p_{|\phi\rangle}(0,0,1\,|\,i-2,i-1,i)$ is equal to $p_{|\phi\rangle}(1\,|\,i)$, since observables $i$ only have two outcomes.}
\end{table}

%%%%%%%%%%%%%%%%%%%%%%%%%%%%%%%%%%%%%%%%%%%%%%%%%%%%%%%%%%%%%%%%%%%

For that, we perform a Bell-CHSH inequality experiment. Its aim is to reach the maximum possible value for $S$ b(see Methods). The results obtained in this experiment are shown in table~\ref{Table1}.
%They are in very good agreement with the maximum value for $S$ predicted by quantum theory.
We also perform a series of additional tests to check that the 8 events in $S$ have the 12 relations of mutual exclusivity shown in Fig.~\ref{Fig1}(b) (see Supplementary Material).

%%%%%%%%%%%%%%%%%%%%%%%%%%%%%%%%%%%%%%%%%%%%%%%%%%%%%%%%%%%%%%%%%%%
%\subsection{The Rome complementary inequality experiment}
%%%%%%%%%%%%%%%%%%%%%%%%%%%%%%%%%%%%%%%%%%%%%%%%%%%%%%%%%%%%%%%%%%%

On the other hand, we perform an independent NC inequality experiment in a different physical system in a distant laboratory. Its aim is to reach the maximum possible value for $R$ (see Methods).
The results obtained in this experiment are shown in table~\ref{Table2}.
We also check that the 8 events in $R$ satisfy the 16 relations of mutual exclusivity in Fig.~\ref{Fig1}(c) and check that the 16 inequalities $W_i \stackrel{\mbox{\tiny{E}}}{\leq} 1$ are satisfied (see Supplementary Material).

%%%%%%%%%%%%%%%%%%%%%%%%%%%%%%%%%%%%%%%%%%%%%%%%%%%%%%%%%%%%%%%%%%%

QT suggests that $S$ can be as high as $2+\sqrt{2}\approx 3.4142$ . If that prediction could genuinely be realized for $S$, then the E principle would restrict $R$ to be upper-bounded by exactly its maximal quantum prediction. This is exactly the meaning of (\ref{result}) and (\ref{Rbound}). When we actually performed the experiment we found $S_{\rm exp}=3.413 \pm 0.013$, which means that nature must restrict
\begin{equation}
 R \stackrel{\mbox{\tiny{E}}}{\leq} 2.344 \pm 0.009,
 \label{Rexpbound}
\end{equation}
per the E principle, this by direct observation instead of {\em assuming} achievability of the quantum maximum prediction for $S$.
%which is in excellent agreement with the maximum predicted by QT, $R=8-4 \sqrt{2} \approx 2.3431$.
This result is a significant improvement over an earlier limit of $R \leq \frac{3 \sqrt{3}}{2} \approx 2.5298$, derived by application of the E principle to a gedankenexperiment consisting of two copies of the scenario graphed in Fig.~\ref{Fig1}(c) \cite{Cabello13}.

Given the obtained experimental value $R_{\rm exp}=2.335 \pm 0.011$ as a lower bound of what nature can reach for $R$, the E principle leads to the conclusion that
\begin{equation}
 S \stackrel{\mbox{\tiny{E}}}{\leq} 3.426 \pm 0.016.
 \label{Sexpbound}
\end{equation}
%which is in excellent agreement with the maximum predicted by QT, $2+\sqrt{2} \approx 3.4142$.
This result similarly improves upon an earlier limit of $S \leq \frac{8}{\sqrt{5}} \approx 3.5777$, derived by application of the E principle to a gedankenexperiment consisting of two copies of the CHSH-Bell scenario graphed in Fig.~\ref{Fig1}(b) \cite{FSABCLA13,Cabello13}.

Summing-up: It is well established, although now superseded, that the E principle provides some form of upper bound when applied to two or more copies of either scenario graphed in Fig.~\ref{Fig1}(b) or (c). However, a result in Ref. \cite{ATC13} implies (\ref{result}), which goes so far as to say that the E principle forbids both upper bounds to be satisfied simultaneously: if the E principle is fundamental, one or both of these upper bounds cannot be experimentally reached. The result of the experiment measuring $R$ puts a tighter upper bound on $S$, and the result of the experiment measuring $S$ puts a tighter upper bound on $R$.

More importantly, the limits on $S$ and $R$ determined in our experiments are not just necessary but also sufficient, that is to say, these bounds derived using the E principle are completely tight. Not only do the bounds (\ref{Rexpbound}) and (\ref{Sexpbound}) together {\em saturate} the inequality (\ref{result}), thus exploiting the full restrictive power of the E principle, but they coincide with the maxima in QT. These are precise limits of nature itself: They can be reached, as we have observed, but they cannot be exceeded, as proven by the E principle.

Result (\ref{Sexpbound}) is the first empirical evidence that the E principle explains the maximum quantum violation of the Bell-CHSH inequality. We relied here on the E principle, together with an observation of nature, to single out the limits of the quantum correlations of the Bell-CHSH equality. We find this to be promising support for the still-unproven conjecture \cite{Cabello13,CDLP13,Yan13,Cabello13b,ATC13} that the E principle, {\em by itself}, can recover this quantum bound. In addition, the fact that even slightly higher values of $S$ would violate the E principle indicates the impossibility of super-quantum correlations satisfying the E principle for the Bell-CHSH inequality experiment.

Similarly, result (\ref{Rexpbound}) is the first empirical evidence that the E principle explains the maximum quantum violation of the NC inequality (\ref{Paris}) and supports the conjecture that the E principle, by itself, singles out the limits of quantum correlations for (\ref{Paris}).

The two specific experimental scenarios we have considered are particularly important, since graphs in Fig.~\ref{Fig1}(b) and (c) are the only simple vertex-transitive graphs for which no previous proof or strong evidence existed that their maximum quantum bound were determined by the E principle \cite{CDLP13,Cabello13c}.

There are logically consistent universes in which the E principle does not hold \cite{Specker60,LSW11} and
theoretical machines that violate the E principle, such as nonlocal boxes \cite{PR94,Cabello13,FSABCLA13}. However, the E principle holds in QT for projective measurements and any generalized measurement in QT is a projective measurement in an enlarged Hilbert space. Previous works have proved that the E principle, by itself, singles out the set of quantum correlations in some scenarios \cite{ATC13} and, at least, the maximum quantum correlations in others \cite{Cabello13,CDLP13,Cabello13c}. Now our experiment shows that, by the mechanism proven in \cite{ATC13}, the E principle determines the observed limits for correlations, in two relevant scenarios for which theoretical proofs were heretofore inconclusive. This suggests that the E principle is a key to a deeper understanding of QT and nature.

The possibility that the E principle is fundamental has fundamental implications for both QT and information theory.

In the E principle - unlike in QT - the basic concept is probability rather than a complex probability amplitude. The E principle looks as an extra axiom to Kolmogorov's axioms of probability theory.
 %rather than an Einstein-type physical principle stating the possibility or impossibility of a certain task.
This extra axiom would be significant if we accept as the zero principle of QT that ``unperformed experiments have no results'' (since results do not correspond to intrinsic properties) \cite{Peres78}. From this follows that it may {\em not} be possible, in general, to measure all observables jointly. Then, the fact that pairwise jointly measurable observables are jointly measurable and, consequently, the E principle is highly nontrivial.
With the E principle, QT appears more like an evolution from Kolmogorov's probability theory rather than a development from Newtonian and Maxwellian physics. The same way Kolmogorov's probability theory is not about coins and dice, QT would not {\em really} be about electrons and photons. This would explain why QT successfully applies to most branches of physics and the flexibility of QT for dealing with new phenomena.

The E principle also provided valuable a-priori restrictions on information processing and communication. For example, non-local boxes, whose information processing capabilities have been extensively investigated \cite{BP05}, turn out to be impossible under the E principle, in the same way as perpetual motion machines are impossible without violating the principles of thermodynamics. Another example is secure communication: if the E principle is a fundamental one, then communication with security guaranteed by fundamental principles would be easier to accomplish. This is so because current security is based on the assumption that the adversary's capabilities are limited by only the impossibility of signalling between causally unconnected parties \cite{SGBMPA06}. The possible eavesdropping strategies that a cryptographer needs to consider are vastly reduced by relying on the E principle, which is notoriously more restrictive.

All this follows from a simple observation: (in QT) pairwise joint measurability implies joint measurability \cite{Specker60}. Unfortunately, the apparent irrelevance of this observation suppressed awareness of its significant implications, until only recently.

%%%%%%%%%%%%%%%%%%%%%%%%%%%%%%%%%%%%%%%%%%%%%%%%%%%%%%%%%%%%%%%%%%%

\section*{Methods}

%%%%%%%%%%%%%%%%%%%%%%%%%%%%%%%%%%%%%%%%%%%%%%%%%%%%%%%%%%%%%%%%%%%

%%%%%%%%%%%%%%%%%%%%%%%%%%%%%%%%%%%%%%%%%%%%%%%%%%%%%%%%%%%%%%%%%%%
% Fig. 2
%%%%%%%%%%%%%%%%%%%%%%%%%%%%%%%%%%%%%%%%%%%%%%%%%%%%%%%%%%%%%%%%%%%

\begin{figure}[t]
\centering
\includegraphics[scale=0.23]{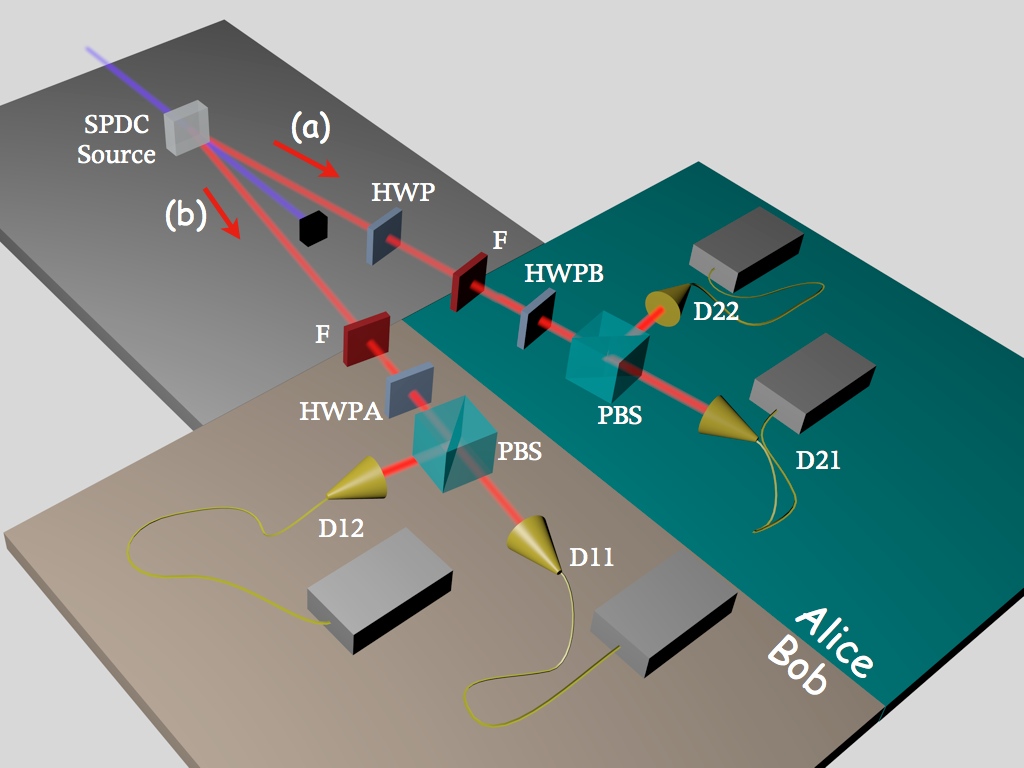}
\caption{\label{Fig2}Setup for the Bell-CHSH inequality experiment. A SPDC source is used to generate a maximally entangled state. To prepare state (\ref{stateS}), a HWP in mode $a$ is set to $-33.75^\circ$. The measurement is performed by HWPA, HWPB plates, PBS and single photon detectors.}
\end{figure}

%%%%%%%%%%%%%%%%%%%%%%%%%%%%%%%%%%%%%%%%%%%%%%%%%%%%%%%%%%%%%%%%%%%

{\bf Setup for the Bell-CHSH inequality experiment}. $S$ is maximized by the following two-qubit state:
\begin{equation}
%\begin{split}
\label{stateS}
|\psi\rangle = \frac{1}{2 \sqrt{2-\sqrt{2}}}\left[|00\rangle-|11\rangle-(1-\sqrt{2}) \left(|01\rangle + |10\rangle\right)\right],
% & = 0.6533 \left(|00\rangle - |11\rangle\right) + 0.2706 \left(|01\rangle + |10\rangle\right),
%\end{split}
\end{equation}
with measurements performed on both qubits in the basis spanned by the Pauli observables, $0= \sigma_z$ and $1= \sigma_x$. The 8 measurement outcomes needed for $S$ are described in table \ref{Table1}. We encode our states in the optical polarizations of pairs of photons as follows:
$|00\rangle=|HH\rangle$, $|01\rangle=|HV\rangle$, $|10\rangle=|VH\rangle$ and $|11\rangle=|VV\rangle$, where $|01\rangle=|HV\rangle$ is the two-photon state for which the first photon is vertically polarized and the second photon is horizontally polarized.

The setup is shown in Fig.~\ref{Fig2}. Ultraviolet light centered at wavelength of 390 nm is focused inside a beta barium borate (BBO) nonlinear crystal to produce photon pairs emitted into two spatial modes (a) and (b) through the degenerate emission of spontaneous parametric down-conversion (SPDC) \cite{Kwiat95}. The source is engineered to prepare the singlet state $|\Psi\rangle = (|HV\rangle - |VH\rangle)/\sqrt{2}$. To obtain the state $|\psi\rangle$ given in (2), we prepare a singlet state and then rotate a half-wave plate (HWP) placed in mode (a) to $-33.75^\circ$ (angle between the waveplate's optical axes direction and horizontal polarization direction).
The polarization measurement is performed using HWPs and polarizing beam splitters (PBSs). At the input of the PBS, we place narrow-bandwidth interference filters (F) ($\delta \lambda = 1$ nm) to guarantee well defined spectral modes. At the output of the PBS, the photons are coupled into 2 m single mode optical fibers followed by actively quenched Si-avalanche photodiodes $D_{ij}$. The measurement time for each pair of local settings is 200 seconds.

%%%%%%%%%%%%%%%%%%%%%%%%%%%%%%%%%%%%%%%%%%%%%%%%%%%%%%%%%%%%%%%%%%%
% Fig. 3
%%%%%%%%%%%%%%%%%%%%%%%%%%%%%%%%%%%%%%%%%%%%%%%%%%%%%%%%%%%%%%%%%%%

\begin{figure}[t]
\centering
\includegraphics[scale=0.23]{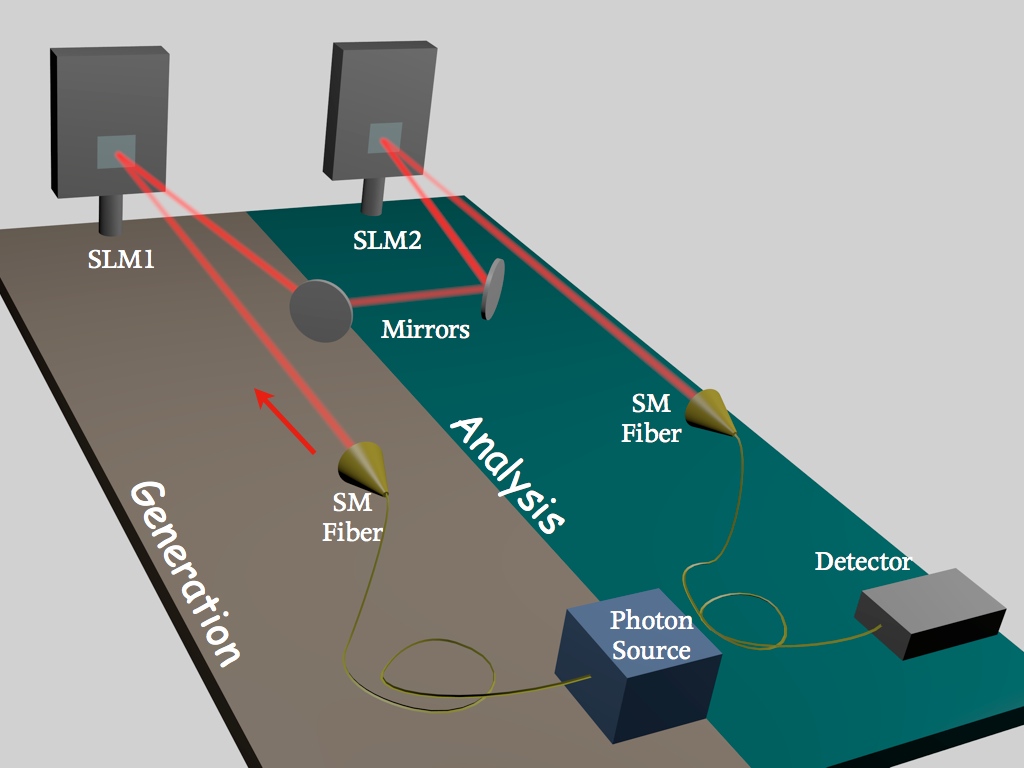}
\caption{\label{Fig3}Setup for the NC inequality experiment. By means of a single mode fiber, the beam from a single photon source is sent to a first SLM (Generation), which provides the states to be measured. Then, projective measurements are performed by means of a second SLM (Analysis), in combination with a single mode fiber and a single photon detector. To avoid Gouy phase shift effect, we have realized an imaging system (not reported in figure) between the screens of the two SLMs.}
\end{figure}

%%%%%%%%%%%%%%%%%%%%%%%%%%%%%%%%%%%%%%%%%%%%%%%%%%%%%%%%%%%%%%%%%%%

{\bf Setup for the NC inequality experiment}. $R$ is maximized by a single five-dimensional five-dimensional quantum system in the state
\begin{equation}
\begin{split}
 | \phi \rangle = &
 \sqrt{1-\frac{1}{\sqrt{2}}} |0\rangle
 +\sqrt{1-\frac{1}{\sqrt{2}}} |1\rangle\\
 &+\sqrt{1-\frac{1}{\sqrt{2}}} |2\rangle
 +\sqrt{\frac{3}{\sqrt{2}}-2} |3\rangle,
\end{split}
\end{equation}
and perform the 8 tests $i = |v_i\rangle \langle v_i|$, with $|v_i\rangle$ defined as follows:
\begin{equation}
\begin{split}
 |v_0\rangle&=|0\rangle,\;\;\;|v_1\rangle=|1\rangle,\;\;\;|v_2\rangle=|2\rangle,\;\\
 |v_3\rangle&=(2-\sqrt{2})|0\rangle+\sqrt{\sqrt{2}-1}|3\rangle-\sqrt{3 \sqrt{2}-4}|4\rangle,\\
 |v_4\rangle&=(3-2 \sqrt{2})|0\rangle+(2-\sqrt{2})|1\rangle+\sqrt{2 \left(5 \sqrt{2}-7\right)}|3\rangle\\
 &+\sqrt{6 \sqrt{2}-8}|4\rangle,\\
 |v_5\rangle&=(2-\sqrt{2})|0\rangle+(3-2 \sqrt{2})|1\rangle+(2-\sqrt{2})|2\rangle\\
 &-2 \sqrt{5 \sqrt{2}-7}|3\rangle,\\
 |v_6\rangle&=(\sqrt{2}-2)|1\rangle+(2 \sqrt{2}-3)|2\rangle-\sqrt{2 \left(5 \sqrt{2}-7\right)}|3\rangle\\
 &+\sqrt{6 \sqrt{2}-8}|4\rangle,\\
 |v_7\rangle&=(\sqrt{2}-2)|2\rangle-\sqrt{\sqrt{2}-1}|3\rangle-\sqrt{3 \sqrt{2}-4}|4\rangle.
\end{split}
 \label{OR}
\end{equation}

We encode the states in photon orbital angular momentum (OAM) space of dimension 5. Each state is a linear combination of the OAM basis states $\{|m\rangle\}^{2}_{m=-2}$.

The experimental setup is illustrated in Fig.~\ref{Fig3}. Single photons in fundamental TEM00 Gaussian state ($m=0$) are prepared in the desired OAM superposition state by means of a spatial light modulator (SLM) (Generation). This device modulates the phase wave front according to computer generated holograms. After the state generation, a second SLM (Analysis) is used in combination with a single mode fiber and single photon detector to perform a projective measurement on the photon state.

This setup allows us to generate and project over all the states needed for the experiment. The generation and measurement processes are completely automatized and computer controlled. We adopt a hologram generation technique that maximizes the fidelity of the states by introducing losses in the beam \cite{DCKNSMS13}. This results in different hologram diffraction efficiencies (see Supplementary Material).

\subsection*{Acknowledgments}

%%%%%%%%%%%%%%%%%%%%%%%%%%%%%%%%%%%%%%%%%%%%%%%%%%%%%%%%%%%%%%%%%%%

The authors thank B. Amaral, M. Ara\'ujo, M. Kleinmann, J. R. Portillo, R. W. Spekkens and M. Terra Cunha for useful conversations, J.-\AA. Larsson, A. J. L\'opez-Tarrida and E. Wolfe for suggestions to improve the manuscript, and E. Wolfe for the figure in the Supplementary Material. This work was supported by Project No.\ FIS2011-29400 (MINECO, Spain) with FEDER funds, the FQXi large grant project ``The Nature of Information in Sequential Quantum Measurements'', the Brazilian program Science without Borders, FIRB Futuro in Ricerca-HYTEQ, the Swedish Research Council (VR), the Linnaeus Center of Excellence ADOPT, the ERC Advanced Grant QOLAPS and Starting Grant 3D-QUEST (3D-Quantum Integrated Optical Simulation; Grant Agreement No. 307783): \url{http://www.3dquest.eu}. M.N. is supported by the international PhD project grant MPD/2009-3/4 from the Foundation for Polish Science.

%%%%%%%%%%%%%%%%%%%%%%%%%%%%%%%%%%%%%%%%%%%%%%%%%%%%%%%%%%%%%%%%%%%

\subsection*{Authors' contributions}

%%%%%%%%%%%%%%%%%%%%%%%%%%%%%%%%%%%%%%%%%%%%%%%%%%%%%%%%%%%%%%%%%%%

M.N. conducted the Bell-CHSH inequality experiment and processed the data assisted by M.B. and A.C., M.N. and E.A. built the two-photon source. F.B. and V.D. conducted the NC inequality experiment and processed the data assisted by F.S. and A.C., V.D. and F.S. built the OAM setup. A.C. conceived the experiment. All authors wrote the paper and agree on its content.

%%%%%%%%%%%%%%%%%%%%%%%%%%%%%%%%%%%%%%%%%%%%%%%%%%%%%%%%%%%%%%%%%%%

\subsection*{Additional information}

%%%%%%%%%%%%%%%%%%%%%%%%%%%%%%%%%%%%%%%%%%%%%%%%%%%%%%%%%%%%%%%%%%%

Correspondence and requests for materials should be addressed to F.S., M.B. or A.C. The authors declare no competing financial interests.

%%%%%%%%%%%%%%%%%%%%%%%%%%%%%%%%%%%%%%%%%%%%%%%%%%%%%%%%%%%%%%%%%%%

\newpage

%%%%%%%%%%%%%%%%%%%%%%%%%%%%%%%%%%%%%%%%%%%%%%%%%%%%%%%%%%%%%%%%%%%

%%%%%%%%%%%%%%%%%%%%%%%%%%%%%%%%%%%%%%%%%%%%%%%%%%%%%%%%%%%%%%%%%%%

\mbox{}
\newpage

\appendix

%%%%%%%%%%%%%%%%%%%%%%%%%%%%%%%%%%%%%%%%%%%%%%%%%%%%%%%%%%%%%%%%%%%

\section{Specker's observation and the exclusivity principle within a general framework of operational theories}

%%%%%%%%%%%%%%%%%%%%%%%%%%%%%%%%%%%%%%%%%%%%%%%%%%%%%%%%%%%%%%%%%%%

We formulate Specker's observation about quantum theory and the E principle within a general framework of operational theories and prove that the E principle is inherently satisfied by any theory in which Specker's observation holds.

%%%%%%%%%%%%%%%%%%%%%%%%%%%%%%%%%%%%%%%%%%%%%%%%%%%%%%%%%%%%%%%%%%%

\subsection{Preparations and tests}

%%%%%%%%%%%%%%%%%%%%%%%%%%%%%%%%%%%%%%%%%%%%%%%%%%%%%%%%%%%%%%%%%%%

Preparations and tests are taken as primitive notions with the following meaning:

A {\em preparation} is a sequence of unambiguous and reproducible experimental procedures.

A {\em test} is a preparation followed by a step in which outcome information is supplied to an observer. This information is not trivial since tests that follow identical preparations may not have identical outcomes.

An {\em operational theory} is one that specifies the probabilities of each possible outcome $X$ of each possible test $M$ given each preparation $P$. We denote these probabilities by $p(X|M;P)$.

The presented framework is independent of the interpretation of probability used; the reader is free to use, e.g., frequentist, propensity, or Bayesian interpretations.

%%%%%%%%%%%%%%%%%%%%%%%%%%%%%%%%%%%%%%%%%%%%%%%%%%%%%%%%%%%%%%%%%%%

\subsection{States and observables}

%%%%%%%%%%%%%%%%%%%%%%%%%%%%%%%%%%%%%%%%%%%%%%%%%%%%%%%%%%%%%%%%%%%

Two preparations are operationally equivalent if they yield identical outcome probability distributions for either test. Each equivalence class of preparations is called a {\em state}. For instance, the state associated with a particular preparation $P_1$ is
\begin{equation}
 \rho_1 \equiv \{P\;|\;\forall M : p(X|M;P) = p(X|M;P_1)\}.
\end{equation}

Two tests are operationally equivalent if they yield identical outcome probability distributions for either preparation. Each equivalence class of tests is called an {\em observable}. For instance, the observable associated with a particular test $M_1$ is
\begin{equation}
 \mu_1 \equiv \{M\;|\;\forall P: p(X|M;P)=p(X|M_1;P)\}.
\end{equation}

%A state can be specified through the list of probabilities for the outcomes of all observables.

%A test may involve a test $M_1$ followed by a second test chosen between $M_2$ or $M_3$, depending on the outcome %of $M_1$.

%%%%%%%%%%%%%%%%%%%%%%%%%%%%%%%%%%%%%%%%%%%%%%%%%%%%%%%%%%%%%%%%%%%

\subsection{Joint measurability of observables}

%%%%%%%%%%%%%%%%%%%%%%%%%%%%%%%%%%%%%%%%%%%%%%%%%%%%%%%%%%%%%%%%%%%

Two observables $\mu_1$ and $\mu_2$ are {\em jointly measurable} if there exists an observable $\mu$ such that: (i) the outcome set of $\mu$, $\sigma(\mu)$, is the Cartesian product of the outcome sets of $\mu_1$ and $\mu_2$, i.e.,
\begin{equation}
 \sigma(\mu)\equiv \{(X_i,X_j)\;|\;X_i \in \sigma(\mu_1), X_j \in \sigma(\mu_2)\},
\end{equation}
and
(ii) for all states $\rho$, the outcome probability distributions for every measurement of $\mu_1$ or $\mu_2$ are recovered as marginals of the outcome probability distribution of $\mu$, i.e.,
\begin{subequations}
\begin{align}
 \forall \rho,\; & \forall X_i \in \sigma(\mu_1) : \nonumber \\
 &p(X_i | \mu_1; \rho) = \sum_{X_j \in \sigma(\mu_2)} p((X_i,X_j) | \mu; \rho),\\
 \forall \rho,\; & \forall X_j \in \sigma(\mu_2) : \nonumber \\
 &p(X_j | \mu_2; \rho) = \sum_{X_i \in \sigma(\mu_1)} p((X_i,X_j) | \mu; \rho). \label{c}
\end{align}
\end{subequations}

$N$ observables $\mu_1,\ldots,\mu_N$ are {\em jointly measurable} if there exists an observable $\mu$ such that: (i') the outcome set of $\mu$ is the Cartesian product of the outcome sets of $\mu_1,\ldots,\mu_N$ and (ii') for all states $\rho$, the outcome probability distributions for every joint measurement of any subset $S \equiv \{\mu_i|i\in I\} \subset \{\mu_1,\ldots,\mu_N\}$, with $I=\{1,\ldots,N\}$, are recovered as marginals of the outcome probability distribution of $\mu$. Denoting by $\mu_S$ an observable associated with a joint measurement of the subset $S$, its outcome set by $\sigma(\mu_S)$ and one of its outcomes by $X_S$, the condition can be expressed as
\begin{equation}
\begin{split}
 \forall S,\; \forall \rho,\; & \forall X_S \in \sigma(\mu_S) : \\
 &p(X_S|\mu_S;\rho) = \sum_{X_t : t \notin I} p((X_1,\ldots,X_N)|\mu;\rho).
\end{split}
\end{equation}

Joint measurability of a set of observables implies pairwise joint measurability of them (i.e., joint measurability of any pair of them). The converse is not necessarily true.

A {\em joint probability distribution} for $N$ observables $\mu_1,\ldots,\mu_N$ exists if, for all subsets $S \equiv \{\mu_i|i\in I\} \subset \{\mu_1,\ldots,\mu_N\}$, with $I=\{1,\ldots,N\}$, for all states $\rho$ and for all $X_S \in \sigma(\mu_S)$, where $\mu_{S}$ ia an observable associated with a joint measurement of $S$, there exists a probability distribution $p(X_1,\ldots,X_N|\rho)$ such that
\begin{equation}
 p(X_S|\mu_S;\rho)=\sum_{X_t : t \notin I} p(X_1,\ldots,X_N|\rho).
\end{equation}

If some observables are jointly measurable then there exists a joint probability distribution for them. The existence of a joint probability distribution for some observables does not imply that they are jointly measurable. The nonexistence of a joint probability distribution for some observables indicates the impossibility of jointly measuring them.

%%%%%%%%%%%%%%%%%%%%%%%%%%%%%%%%%%%%%%%%%%%%%%%%%%%%%%%%%%%%%%%%%%%

\subsection{Events}

%%%%%%%%%%%%%%%%%%%%%%%%%%%%%%%%%%%%%%%%%%%%%%%%%%%%%%%%%%%%%%%%%%%

We are interested in a specific type of preparations: those resulting from a test $M$ with outcome $X$ on a previous preparation $P$. We denote these preparations by $P' \equiv X|M;P$.

Two of these preparations are operationally equivalent if they yield identical outcome probability distributions for either subsequent test $M'$. Each equivalence class of these preparations is called an {\em event}. For instance, the event associated with a particular preparation $P'_1 \equiv X_1|M_1;P_1$ is
%\begin{widetext}
\begin{equation}
 \epsilon_1 \equiv \left\{P'\;|\;\forall M' : p(X'|M';P') = p(X'|M';P'_1)\right\}.
\end{equation}
%\end{widetext}
Notice that the term ``event'', which is usually restricted to designate the {\em outcome} $X$ of test $M$ on preparation $P$, here designates the {\em state} after test $M$ with outcome $X$ on preparation $P$. The probability of an event is therefore the probability of {\em transforming} one state (e.g., the one associated with $P_1$) into another (e.g., the one associated with $P'_1$). In a given non-contextuality (NC) inequality, all probabilities (of events) are probabilities of different transformations of the {\em same} state.

%%%%%%%%%%%%%%%%%%%%%%%%%%%%%%%%%%%%%%%%%%%%%%%%%%%%%%%%%%%%%%%%%%%

\subsection{Mutual exclusivity of events}

%%%%%%%%%%%%%%%%%%%%%%%%%%%%%%%%%%%%%%%%%%%%%%%%%%%%%%%%%%%%%%%%%%%

Two events $\epsilon_1$ and $\epsilon_2$ are {\em mutually exclusive} if there exist two jointly measurable observables $\mu_1$, univocally defined by $\epsilon_1$, and $\mu_2$, univocally defined by $\epsilon_2$, that distinguish between them, i.e., if there exists an observable $\mu$ associated with a joint measurement of $\mu_1$ and $\mu_2$ such that there are $X_1 \subset \sigma(\mu)$ and $X_2 \subset \sigma(\mu)$ with $X_1 \cap X_2 = \emptyset$ such that
\begin{subequations}
\begin{align}
 &p(X_1 | \mu; \epsilon_1) = 1, \label{condition}\\
 &p(X_2 | \mu; \epsilon_2) = 1. \label{conditionb}
\end{align}
\end{subequations}

The $N$ events of a set ${\cal E}=\{\epsilon_1,\ldots,\epsilon_N\}$ are {\em jointly exclusive} if there exists a set of $N$ jointly measurable observables ${\cal M}=\{\mu_1,\ldots,\mu_N\}$ that distinguish between the events in any subset of ${\cal E}$.

Joint exclusivity of a set of events implies mutual exclusivity of any pair of them. The converse is not necessarily true.

%%%%%%%%%%%%%%%%%%%%%%%%%%%%%%%%%%%%%%%%%%%%%%%%%%%%%%%%%%%%%%%%%%%

\subsection{Specker's observation and the E principle}

%%%%%%%%%%%%%%%%%%%%%%%%%%%%%%%%%%%%%%%%%%%%%%%%%%%%%%%%%%%%%%%%%%%

{\em Specker's observation.} Specker pointed out that, in quantum theory, pairwise joint measurability of a set ${\cal M}$ of observables implies joint measurability of ${\cal M}$, while in other theories this implication does not need to hold \cite{Specker60}. Later, Specker conjectured that this is ``the fundamental theorem'' of quantum theory (see \url{http://vimeo.com/52923835}).

{\em The E principle} states that any set of pairwise mutually exclusive events is jointly exclusive. Therefore, from Kolmogorov's axioms of probability, the sum of their probabilities cannot be higher than~1.

{\em Lemma:} In any theory in which pairwise joint measurability of observables implies joint measurability of observables, pairwise mutual exclusivity of events implies joint exclusivity of events.

{\em Proof:} If the events in a set ${\cal E}=\{\epsilon_1,\ldots,\epsilon_N\}$ are pairwise exclusive, there exists a set ${\cal M}=\{\mu_1,\ldots,\mu_N\}$ of pairwise jointly measurable observables that permits to distinguish between any two events in ${\cal E}$. If pairwise joint measurability of ${\cal M}$ implies joint measurability of ${\cal M}$, then ${\cal M}$ permits to distinguish between the events in any subset of ${\cal E}$.\hfill \endproof

The converse implication, namely, that in any theory in which pairwise mutual exclusivity implies joint exclusivity also pairwise joint measurability implies joint measurability, is not necessarily true.

%%%%%%%%%%%%%%%%%%%%%%%%%%%%%%%%%%%%%%%%%%%%%%%%%%%%%%%%%%%%%%%%%%%

\section{Experimental test of the relations of mutual exclusivity}

%%%%%%%%%%%%%%%%%%%%%%%%%%%%%%%%%%%%%%%%%%%%%%%%%%%%%%%%%%%%%%%%%%%

As a complement to the tests of the Bell-CHSH and NC inequalities, we make several tests to check that the 8 events whose probabilities are tested in the Bell-CHSH inequality experiment satisfy the 12 relations of mutual exclusivity represented in Fig.~\ref{Fig1}(b) and the 8 events whose probabilities are tested in the NC inequality satisfy the 16 relations of mutual exclusivity represented in Fig.~\ref{Fig1}(c).

%%%%%%%%%%%%%%%%%%%%%%%%%%%%%%%%%%%%%%%%%%%%%%%%%%%%%%%%%%%%%%%%%%%

\subsection{Mutual exclusivity between the events in the Bell-CHSH inequality}

%%%%%%%%%%%%%%%%%%%%%%%%%%%%%%%%%%%%%%%%%%%%%%%%%%%%%%%%%%%%%%%%%%%

Two events are mutually exclusive if they correspond to different outcomes of an observable $\mu$. We identify a test defining $\mu$ for each pair of events. There are two cases:

4 of the relations of mutual exclusivity occur between events in which Alice implements $\mu_A$ and Bob implements $\mu_B$. For them, $\mu$ is simply an observable associated with a joint measurement of $\mu_A$ and $\mu_B$.

The other 8 relations of mutual exclusivity occur between events in which one of the observers, e.g. Alice, implements $\mu_A$ while the other observer implements $\mu_B$ or $\mu_{B'}$ depending on the outcome of $\mu_A$. This gives a test defining $\mu$, since $\mu_A$ and $\mu_B$ are jointly measurable and $\mu_A$ and $\mu_{B'}$ are jointly measurable.

As an additional test, we experimentally show that conditions (\ref{condition}) and (\ref{conditionb}) are satisfied for any pair of mutually exclusive events. This is shown in table~\ref{Table4}.

%%%%%%%%%%%%%%%%%%%%%%%%%%%%%%%%%%%%%%%%%%%%%%%%%%%%%%%%%%%%%%%%%%%
% Table 4
%%%%%%%%%%%%%%%%%%%%%%%%%%%%%%%%%%%%%%%%%%%%%%%%%%%%%%%%%%%%%%%%%%%

\begin{table}[ht]
\centering
\begin{tabular}{c c}
\hline \hline
Probability & Experimental value \\ [0.5ex]
\hline
$p(1|\mu_0;u_0)$ & $0.997916 \pm 0.000076$ \\
$p(1|\mu_4;u_0)$ & $1.8 \times 10^{-6} \pm 1.9 \times 10^{-6}$ \\
$p(1|\mu_3;u_0)$ & $0.000635 \pm 0.000030$ \\
$p(1|\mu_5;u_0)$ & $0.000373 \pm 0.000024$ \\
$p(1|\mu_3;u_3)$ & $0.997823 \pm 0.000059$ \\
$p(1|\mu_7;u_3)$ & $2.2 \times 10^{-6} \pm 1.9 \times 10^{-6}$ \\
$p(1|\mu_6;u_3)$ & $0.000639 \pm 0.000040$ \\
$p(1|\mu_0;u_3)$ & $0.000437 \pm 0.000017$ \\
$p(1|\mu_6;u_6)$ & $0.997484 \pm 0.000062$ \\
$p(1|\mu_2;u_6)$ & $3.3 \times 10^{-6} \pm 2.5 \times 10^{-6}$ \\
$p(1|\mu_3;u_6)$ & $0.000651 \pm 0.000037$ \\
$p(1|\mu_1;u_6)$ & $0.000606 \pm 0.000032$ \\
$p(1|\mu_1;u_1)$ & $0.996928 \pm 0.000096$ \\
$p(1|\mu_5;u_1)$ & $1.1 \times 10^{-6} \pm 1.2 \times10^{-6}$ \\
$p(1|\mu_6;u_1)$ & $0.000666 \pm 0.000045$ \\
$p(1|\mu_4;u_1)$ & $0.000937 \pm 0.000040$ \\
$p(1|\mu_4;u_4)$ & $0.995531 \pm 0.000075$ \\
$p(1|\mu_0;u_4)$ & $6.4 \times 10^{-6} \pm 2.7 \times 10^{-6}$ \\
$p(1|\mu_1;u_4)$ & $0.001419 \pm 0.000040$ \\
$p(1|\mu_7;u_4)$ & $0.000864 \pm 0.000027$ \\
$p(1|\mu_7;u_7)$ & $0.992080 \pm 0.000120$ \\
$p(1|\mu_3;u_7)$ & $1.49 \times 10^{-5} \pm 5.8 \times 10^{-6}$ \\
$p(1|\mu_4;u_7)$ & $0.001816 \pm 0.000047$ \\
$p(1|\mu_2;u_7)$ & $0.000637 \pm 0.000027$ \\
$p(1|\mu_2;u_2)$ & $0.995735 \pm 0.000098$ \\
$p(1|\mu_6;u_2)$ & $3.3 \times 10^{-6} \pm 2.1 \times 10^{-6}$ \\
$p(1|\mu_7;u_2)$ & $0.000477 \pm 0.000022$ \\
$p(1|\mu_5;u_2)$ & $0.001469 \pm 0.000026$ \\
$p(1|\mu_5;u_5)$ & $0.996841 \pm 0.000066$ \\
$p(1|\mu_1;u_5)$ & $2.9 \times 10^{-6} \pm 2.1 \times 10^{-6}$ \\
$p(1|\mu_2;u_5)$ & $0.000893 \pm 0.000041$ \\
$p(1|\mu_0;u_5)$ & $0.001314 \pm 0.000051$ \\
\hline
\hline
\end{tabular}
\caption{\label{Table4}Results of the tests to check that the events in the Bell-CHSH inequality experiment satisfy the relations of exclusivity in Fig.~\ref{Fig1}(b). $p(1|\mu_j;u_i)$ denotes the probability of obtaining the result~1 when the observable $\mu_j$ (which corresponds in quantum theory to $|u_j\rangle \langle u_j|$) is measured on the state $|u_i\rangle$.}
\end{table}

%%%%%%%%%%%%%%%%%%%%%%%%%%%%%%%%%%%%%%%%%%%%%%%%%%%%%%%%%%%%%%%%%%%

\subsection{Mutual exclusivity between the events in the NC inequality}

%%%%%%%%%%%%%%%%%%%%%%%%%%%%%%%%%%%%%%%%%%%%%%%%%%%%%%%%%%%%%%%%%%%

Two events are mutually exclusive if they correspond to different outcomes of an observable $\mu$. We identify a test defining $\mu$ for each pair of events. For that, we prepare 16 additional states $| w_i \rangle$ that allows us to define an 5-outcome observable for each of the 8 triangles in Fig.~\ref{Fig1}(b). These states are specified in table~\ref{Table5}. Each of these 5-outcome observables distinguishes between each pair of events in the corresponding triangle.

As an additional test, we experimentally show that conditions (\ref{condition}) and (\ref{conditionb}) are satisfied for any pair of mutually exclusive events. For this, we measure the fidelity of each state $| v_i \rangle$, defined by $p(1|\mu'_i;v_i)$ and the probabilities $p(1|\mu'_i;v_j)$ where $\mu'_i$ is the observable with outcome set $\{0,1\}$ represented in quantum theory by $|v_i \rangle \langle v_i|$. The experimental results are shown in tables~\ref{Table5} and \ref{Table6}.

%%%%%%%%%%%%%%%%%%%%%%%%%%%%%%%%%%%%%%%%%%%%%%%%%%%%%%%%%%%%%%%%%%%
% Table 5
%%%%%%%%%%%%%%%%%%%%%%%%%%%%%%%%%%%%%%%%%%%%%%%%%%%%%%%%%%%%%%%%%%%

\begin{table*}[htbp]
\centering
\small
{\footnotesize
\begin{tabular}{lllll}
\hline \hline \noalign{\vskip 1mm}
Basis 				 & State 	    & Components 						  	 & Fidelity 			& Efficiency	  \\
\noalign{\vskip 1mm}
\hline
\noalign{\vskip 2mm}
\multirow{5}*{I} & $\ket{v_0}$ & (1,0,0,0,0)							 & $(98.5\pm0.2)\%$ 	& $(98\pm1)\%$   \\
			    	  & $\ket{v_6}$ & $(0, -0.586, -0.172, -0.377, 0.697)$   & $(98.2\pm0.2)\%$ 	& $(42.6\pm0.8)\%$\\
			    	  & $\ket{v_7}$ & $(0, 0, -0.586, -0.644, -0.493)$		 & $(99.3\pm0.1)\%$ 	& $(33.2\pm0.8)\%$\\
				   	  & $\ket{w_1}$ & $(0, -0.806, 0.206, 0.206, -0.515)$ 	 & $(99.2\pm0.1)\%$ 	& $(39.5\pm0.8)\%$\\
					  & $\ket{w_2}$ & $ (0, 0.086, 0.76, -0.63, -0.082)$ 	 & $(98.2\pm0.2)\%$ 	& $(62.3\pm0.8)\%$\\
\noalign{\vskip 1mm}
\hline
\noalign{\vskip 1mm}
\multirow{5}*{II} & $\ket{v_0}$ & (1,0,0,0,0) 						  	 & $(99.6\pm0.1)\%$ 	& $(98\pm1)\%$	  \\
			    	  & $\ket{v_1}$ & (0,1,0,0,0) 							 & $(98.9\pm0.2)\%$		& $(97\pm1)\%$	  \\
			    	  & $\ket{v_2}$ & (0,0,1,0,0)							 & $(98.0\pm0.2)\%$		& $(100\pm1)\%$	  \\
					  & $\ket{w_3}$ & $(0,0,0,0.707,0.707)$					 & $(98.9\pm0.2)\%$		& $(47.0\pm0.8)\%$\\
				  	  & $\ket{w_4}$ & $(0,0,0,0.707,-0.707)$ 				 & $(99.5\pm0.1)\%$		& $(45.4\pm0.8)\%$\\
\noalign{\vskip 1mm}
\hline
\noalign{\vskip 1mm}
\multirow{5}*{III} & $\ket{v_3}$ & $(0.586, 0, 0, 0.644, -0.493)$ 		 & $(99.0\pm0.2)\%$		& $(36.0\pm0.8)\%$\\
					  & $\ket{v_4}$ & $(0.172, 0.586, 0, 0.377, 0.697)$ 	 & $(98.9\pm0.2)\%$		& $(29.1\pm0.8)\%$\\
				 	  & $\ket{v_5}$ & $(0.586, 0.172, 0.586, -0.533, 0)$  	 & $(98.7\pm0.2)\%$		& $(38.1\pm0.8)\%$\\
					  & $\ket{w_5}$ & $(0.202, -0.787, 0.213, 0.202, 0.503)$ & $(98.0\pm0.2)\% $	& $(42.7\pm0.8)\%$\\
					  &$ \ket{w_6}$ & $(0.494, -0.0855, -0.782, -0.345, 0.137)$ & $(99.5\pm0.1)\%$  & $(40.7\pm0.8)\%$\\
\noalign{\vskip 1mm}
\hline
\noalign{\vskip 1mm}
\multirow{5}*{IV} & $\ket{v_0}$ & $(1,0,0,0,0)$ 						 & $(99.4\pm0.1)\%$		& $(98\pm1)\%$\\
					  & $\ket{v_1}$ & $(0,1,0,0,0)$ 						 & $(99.5\pm0.1)\%$		& $(97\pm1)\%$\\
				 	  & $\ket{v_7}$ & $(0, 0, -0.586, -0.644, -0.493)$  	 & $(99.6\pm0.1)\%$		& $(33.2\pm0.8)\%$\\
					  & $\ket{w_7}$ & $(0.202, -0.787, 0.213, 0.202, 0.503)$ & $(99.3\pm0.1)\% $	& $(55.5\pm0.8)\%$\\
					  &$ \ket{w_8}$ & $(0.494, -0.0855, -0.782, -0.345, 0.137)$ & $(98.2\pm0.2)\%$  & $(45.1\pm0.8)\%$\\
\noalign{\vskip 1mm}
\hline
\noalign{\vskip 1mm}
\multirow{5}*{V} & $\ket{v_1}$ & $(0,1,0,0,0)$ 						 & $(99.5\pm0.1)\%$		& $(97\pm1)\%$\\
					  & $\ket{v_2}$ & $(0,0,1,0,0)$ 						 & $(98.8\pm0.2)\%$		& $(100\pm1)\%$\\
				 	  & $\ket{v_3}$ & $(0.586, 0, 0, 0.644, -0.493)$  		 & $(98.7\pm0.2)\%$		& $(36.0\pm0.8)\%$\\
					  & $\ket{w_9}$ & $(0.202, -0.787, 0.213, 0.202, 0.503)$ & $(99.5\pm0.1)\% $	& $(37.2\pm0.8)\%$\\
					  &$ \ket{w_{10}}$ & $(0.494, -0.0855, -0.782, -0.345, 0.137)$ & $(99.1\pm0.1)\%$  & $(42.1\pm0.8)\%$\\
\noalign{\vskip 1mm}
\hline
\noalign{\vskip 1mm}
\multirow{5}*{VI} & $\ket{v_2}$ & $(0,0,1,0,0)$ 						 & $(98.5\pm0.2)\%$		& $(100\pm1)\%$\\
					  & $\ket{v_3}$ & $(0.586, 0, 0, 0.644, -0.493)$ 		 & $(99.2\pm0.1)\%$		& $(36.0\pm0.8)\%$\\
				 	  & $\ket{v_4}$ & $(0.172, 0.586, 0, 0.377, 0.697)$  	 & $(99.6\pm0.1)\%$		& $(29.1\pm0.8)\%$\\
					  & $\ket{w_{11}}$ & $(0.202, -0.787, 0.213, 0.202, 0.503)$ & $(98.9\pm0.2)\% $	& $(44.8\pm0.8)\%$\\
					  &$ \ket{w_{12}}$ & $(0.494, -0.0855, -0.782, -0.345, 0.137)$ & $(99.1\pm0.1)\%$  & $(40.5\pm0.8)\%$\\
\noalign{\vskip 1mm}
\hline
\noalign{\vskip 1mm}
\multirow{5}*{VII} & $\ket{v_4}$ & $(0.172, 0.586, 0, 0.377, 0.697)$ 	 & $(98.9\pm0.2)\%$		& $(29.1\pm0.8)\%$\\
					  & $\ket{v_5}$ & $(0.586, 0.172, 0.586, -0.533, 0)$ 	 & $(98.7\pm0.2)\%$		& $(38.1\pm0.8)\%$\\
				 	  & $\ket{v_6}$ & $(0, -0.586, -0.172, -0.377, 0.697)$   & $(99.2\pm0.1)\%$		& $(40.7\pm0.8)\%$\\
					  & $\ket{w_{13}}$ & $(0.202, -0.787, 0.213, 0.202, 0.503)$ & $(99.5\pm0.1)\% $	& $(46.7\pm0.8)\%$\\
					  &$ \ket{w_{14}}$ & $(0.494, -0.0855, -0.782, -0.345, 0.137)$ & $(99.2\pm0.1)\%$  & $(36.5\pm0.8)\%$\\
\noalign{\vskip 1mm}
\hline
\noalign{\vskip 1mm}
\multirow{5}*{VIII} & $\ket{v_5}$ & $(0.586, 0.172, 0.586, -0.533, 0)$ 	 & $(98.5\pm0.2)\%$		& $(38.1\pm0.8)\%$\\
					  & $\ket{v_6}$ & $(0, -0.586, -0.172, -0.377, 0.697)$ 	 & $(98.6\pm0.2)\%$		& $(40.7\pm0.8)\%$\\
				 	  & $\ket{v_7}$ & $(0, 0, -0.586, -0.644, -0.493)$  	 & $(100.0\pm0.1)\%$		& $(33.2\pm0.8)\%$\\
					  & $\ket{w_{15}}$ & $(0.202, -0.787, 0.213, 0.202, 0.503)$ & $(99.3\pm0.2)\% $	& $(36.1\pm0.8)\%$\\
					  &$ \ket{w_{16}}$ & $(0.494, -0.0855, -0.782, -0.345, 0.137)$ & $(99.6\pm0.1)\%$  & $(39.5\pm0.8)\%$\\
\noalign{\vskip 2mm}
\hline \hline
\end{tabular}
}
\caption{\label{Table5}Complete measurement bases for the NC inequality experiment. The table reports, for each state, both the classical fidelity and the relative diffraction efficiency with respect to the state $|v_2\rangle$, corresponding to TEM00 Gaussian state. Mean base fidelities are $(98.7\pm0.2)\%$, $(99.0\pm0.1)\%$, $(98.8\pm0.1)\%$, $(99.2\pm0.1)\%$, $(99.1\pm0.1)\%$, $(99.1\pm0.1)\%$, $(99.1\pm0.1)\%$ and $(99.2\pm0.1)\%$, for base 1, 2, 3, 4, 5, 6, 7 and 8 respectively. The mean fidelity value is equal to $(99.0\pm0.1)\%$.}
\end{table*}

%%%%%%%%%%%%%%%%%%%%%%%%%%%%%%%%%%%%%%%%%%%%%%%%%%%%%%%%%%%%%%%%%%%
% Table 6
%%%%%%%%%%%%%%%%%%%%%%%%%%%%%%%%%%%%%%%%%%%%%%%%%%%%%%%%%%%%%%%%%%%

\begin{table}[ht]
\centering
\begin{tabular}{c c}
\hline \hline
Probability & Experimental value \\ [0.5ex]
\hline
$p(1|\mu'_0;v_0)$ &  $0.9920 \pm 0.0010$ \\
$p(1|\mu'_1;v_0)$ &  $0.0026 \pm 0.0007$ \\
$p(1|\mu'_2;v_0)$ &  $0.0026 \pm 0.0007$ \\
$p(1|\mu'_6;v_0)$ &  $0.0014 \pm 0.0005$ \\
$p(1|\mu'_7;v_0)$ &  $0.0012 \pm 0.0005$ \\
$p(1|\mu'_1;v_1)$ &  $0.9930 \pm 0.0010$ \\
$p(1|\mu'_0;v_1)$ &  $0.0022 \pm 0.0007$ \\
$p(1|\mu'_2;v_1)$ &  $0.0040 \pm 0.0009$ \\
$p(1|\mu'_3;v_1)$ &  $0.0007 \pm 0.0004$ \\
$p(1|\mu'_7;v_1)$ &  $0.0029 \pm 0.0008$ \\
$p(1|\mu'_2;v_2)$ &  $0.9840 \pm 0.0010$ \\
$p(1|\mu'_0;v_2)$ &  $0.0018 \pm 0.0006$ \\
$p(1|\mu'_1;v_2)$ &  $0.0019 \pm 0.0006$ \\
$p(1|\mu'_3;v_2)$ &  $0.0033 \pm 0.0008$ \\
$p(1|\mu'_4;v_2)$ &  $0.007 \pm 0.001$ \\
$p(1|\mu'_3;v_3)$ &  $0.9890 \pm 0.0010$ \\
$p(1|\mu'_1;v_3)$ &  $0.0040 \pm 0.0009$ \\
$p(1|\mu'_2;v_3)$ &  $0.0026 \pm 0.0007$ \\
$p(1|\mu'_4;v_3)$ &  $0.006 \pm 0.001$ \\
$p(1|\mu'_5;v_3)$ &  $0.0015 \pm 0.0005$ \\
$p(1|\mu'_4;v_4)$ &  $0.9910 \pm 0.0010$ \\
$p(1|\mu'_2;v_4)$ &  $0.0008 \pm 0.0004$ \\
$p(1|\mu'_3;v_4)$ &  $0.0019 \pm 0.0006$ \\
$p(1|\mu'_5;v_4)$ &  $0.0009 \pm 0.0005$ \\
$p(1|\mu'_6;v_4)$ &  $0.008 \pm 0.001$ \\
$p(1|\mu'_5;v_5)$ &  $0.9860 \pm 0.0010$ \\
$p(1|\mu'_3;v_5)$ &  $0.0030 \pm 0.0008$ \\
$p(1|\mu'_4;v_5)$ &  $0.0018 \pm 0.0006$ \\
$p(1|\mu'_6;v_5)$ &  $0.007 \pm 0.001$ \\
$p(1|\mu'_7;v_5)$ &  $0.0007 \pm 0.0004$ \\
$p(1|\mu'_6;v_6)$ &  $0.9870 \pm 0.0010$ \\
$p(1|\mu'_0;v_6)$ &  $0.0028 \pm 0.0008$ \\
$p(1|\mu'_4;v_6)$ &  $0.0007 \pm 0.0009$ \\
$p(1|\mu'_5;v_6)$ &  $0.0041 \pm 0.0009$ \\
$p(1|\mu'_7;v_6)$ &  $0.0014 \pm 0.0005$ \\
$p(1|\mu'_7;v_7)$ &  $0.9960 \pm 0.0010$ \\
$p(1|\mu'_0;v_7)$ &  $0.006 \pm 0.001$ \\
$p(1|\mu'_1;v_7)$ &  $0.0029 \pm 0.0008$ \\
$p(1|\mu'_5;v_7)$ &  $0.0009 \pm 0.0005$ \\
$p(1|\mu'_6;v_7)$ &  $0.0043 \pm 0.0009$ \\
 \hline \hline
\end{tabular}
\caption{\label{Table6}Results of the tests to check that the events in the NC inequality experiment satisfy the relations of exclusivity in Fig.~\ref{Fig1}(c). $p(1|\mu'_j;v_i)$ denotes the probability of obtaining the result~1 when the observable $\mu'_j$  (which corresponds in quantum theory to $|v_j\rangle \langle v_j|$) is measured on the state $|v_i\rangle$.}
\end{table}

%%%%%%%%%%%%%%%%%%%%%%%%%%%%%%%%%%%%%%%%%%%%%%%%%%%%%%%%%%%%%%%%%%%

\section{Exclusivity inequalities}

%%%%%%%%%%%%%%%%%%%%%%%%%%%%%%%%%%%%%%%%%%%%%%%%%%%%%%%%%%%%%%%%%%%

We use the data in tables~\ref{Table1} and \ref{Table2} to check that the 16 exclusivity inequalities $W_i \stackrel{\mbox{\tiny{E}}}{\leq} 1$, with $i=1,\ldots,16$, are satisfied in our experiment. The results are in table~\ref{Table3}.

%%%%%%%%%%%%%%%%%%%%%%%%%%%%%%%%%%%%%%%%%%%%%%%%%%%%%%%%%%%%%%%%%%%
% Table 3
%%%%%%%%%%%%%%%%%%%%%%%%%%%%%%%%%%%%%%%%%%%%%%%%%%%%%%%%%%%%%%%%%%%
\begin{table}[h]
\begin{tabular}{cccccccccc}
\hline
\hline
 & $u_0$ & $u_1$ & $u_2$ & $u_3$ & $u_4$ & $u_5$ & $u_6$ & $u_7$ & Experimental value \\ [0.5ex]
\hline
$W_{1}$ & $v_0$ & $v_1$ & $v_2$ & $v_3$ & $v_4$ & $v_5$ & $v_6$ & $v_7$ & $0.997\pm0.016$ \\
$W_{2}$ & $v_1$ & $v_2$ & $v_3$ & $v_4$ & $v_5$ & $v_6$ & $v_7$ & $v_0$ & $0.997\pm0.016$ \\
$W_{3}$ & $v_2$ & $v_3$ & $v_4$ & $v_5$ & $v_6$ & $v_7$ & $v_0$ & $v_1$ & $0.996\pm0.016$ \\
$W_{4}$ & $v_3$ & $v_4$ & $v_5$ & $v_6$ & $v_7$ & $v_0$ & $v_1$ & $v_2$ & $0.996\pm0.016$ \\
$W_{5}$ & $v_4$ & $v_5$ & $v_6$ & $v_7$ & $v_0$ & $v_1$ & $v_2$ & $v_3$ & $0.996\pm0.016$ \\
$W_{6}$ & $v_5$ & $v_6$ & $v_7$ & $v_0$ & $v_1$ & $v_2$ & $v_3$ & $v_4$ & $0.996\pm0.016$ \\
$W_{7}$ & $v_6$ & $v_7$ & $v_0$ & $v_1$ & $v_2$ & $v_3$ & $v_4$ & $v_5$ & $0.996\pm0.016$ \\
$W_{8}$ & $v_7$ & $v_0$ & $v_1$ & $v_2$ & $v_3$ & $v_4$ & $v_5$ & $v_6$ & $0.996\pm0.016$ \\
$W_{9}$ & $v_0$ & $v_7$ & $v_6$ & $v_5$ & $v_4$ & $v_3$ & $v_2$ & $v_1$ & $0.996\pm0.016$ \\
$W_{10}$& $v_1$ & $v_0$ & $v_7$ & $v_6$ & $v_5$ & $v_4$ & $v_3$ & $v_2$ & $0.996\pm0.016$ \\
$W_{11}$& $v_2$ & $v_1$ & $v_0$ & $v_7$ & $v_6$ & $v_5$ & $v_4$ & $v_3$ & $0.996\pm0.016$ \\
$W_{12}$& $v_3$ & $v_2$ & $v_1$ & $v_0$ & $v_7$ & $v_6$ & $v_5$ & $v_4$ & $0.997\pm0.016$ \\
$W_{13}$& $v_4$ & $v_3$ & $v_2$ & $v_1$ & $v_0$ & $v_7$ & $v_6$ & $v_5$ & $0.996\pm0.016$ \\
$W_{14}$& $v_5$ & $v_4$ & $v_3$ & $v_2$ & $v_1$ & $v_0$ & $v_7$ & $v_6$ & $0.996\pm0.016$ \\
$W_{15}$& $v_6$ & $v_5$ & $v_4$ & $v_3$ & $v_2$ & $v_1$ & $v_0$ & $v_7$ & $0.996\pm0.016$ \\
$W_{16}$& $v_7$ & $v_6$ & $v_5$ & $v_4$ & $v_3$ & $v_2$ & $v_1$ & $v_0$ & $0.996\pm0.016$ \\
\hline
\hline
\end{tabular}
\caption{\label{Table3}Experimental results of the tests of the 16 exclusivity inequalities. The 8 global events $(u_i,v_j)$ in each $W_k$ are given by the table by combining event $u_i$ of the first row with event $v_j$ in the intersection between $W_k$'s row and $u_i$'s column. For example, $W_{1}$ is also defined in (\ref{W1}).}
\end{table}

%%%%%%%%%%%%%%%%%%%%%%%%%%%%%%%%%%%%%%%%%%%%%%%%%%%%%%%%%%%%%%%%%%%

The fact that we observe an experimental value compatible with 1 for each of the 16 inequalities indicates that the experimental results for $S$ and $R$ are both in the limit allowed by the E principle.

%%%%%%%%%%%%%%%%%%%%%%%%%%%%%%%%%%%%%%%%%%%%%%%%%%%%%%%%%%%%%%%%%%%

\section{Exclusivity graph of the complete two-city experiment}

%%%%%%%%%%%%%%%%%%%%%%%%%%%%%%%%%%%%%%%%%%%%%%%%%%%%%%%%%%%%%%%%%%%

\begin{figure*}[t]
\centering
\includegraphics[scale=0.68]{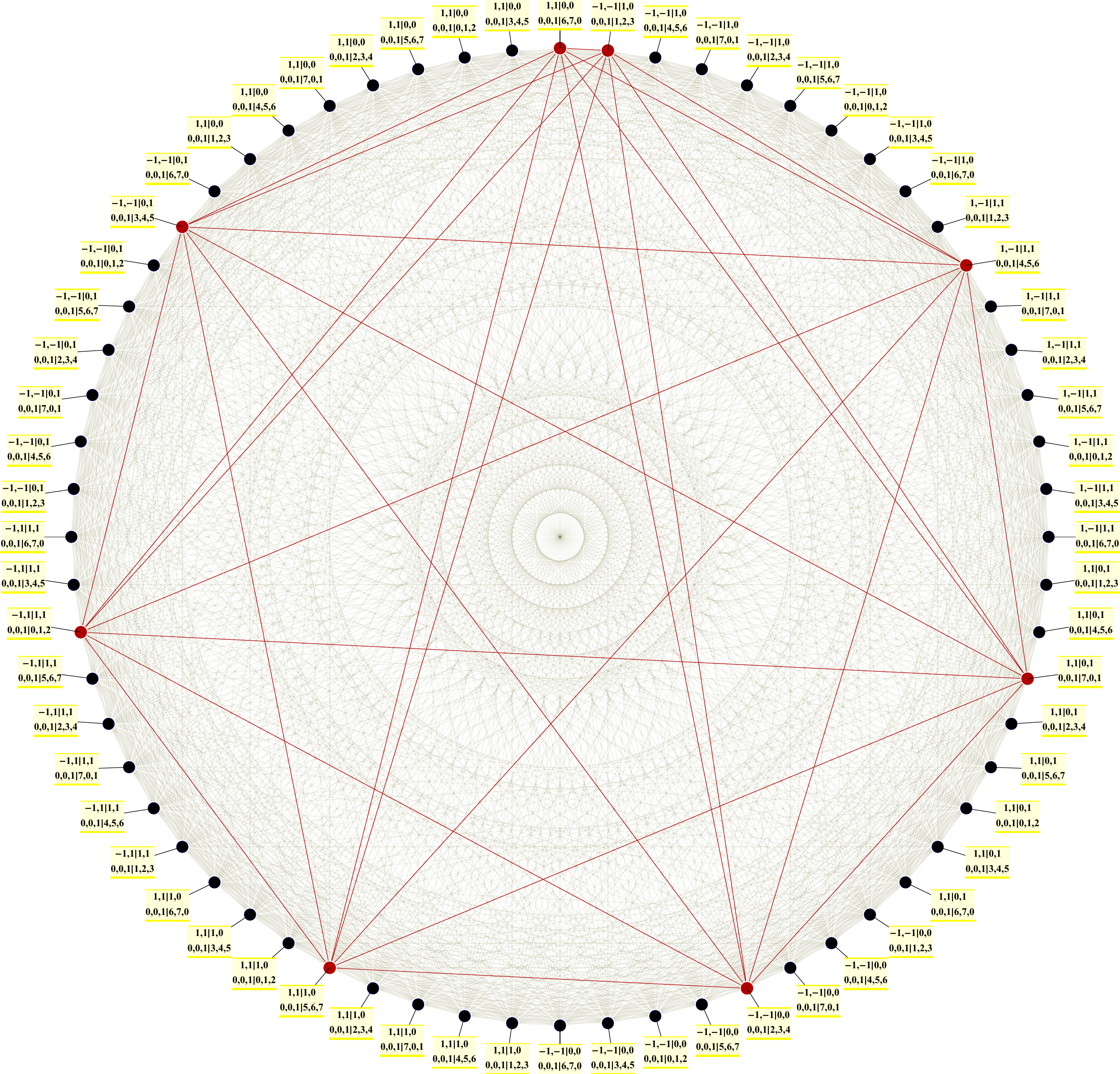}
\vspace{5mm}
\caption{\label{Fig4}Exclusivity graph of the two-city experiment. The 64 dots represent the 64 global events $(u_i,v_j)$. Two dots are connected by an edge if the corresponding events are mutually exclusive. Each $(u_i,v_j)$ is indicated by providing the explicit expression of $u_i$ (for instance, $u_0=1,1|0,0$; see table~\ref{Table1}) followed by the one of $v_j$ (see table~\ref{Table2}). The 8-vertex subgraph with red edges corresponds to the correlations in $W_9$ (see table~\ref{Table3}). Figure courtesy of Elie Wolfe.}
\end{figure*}

Fig.~\ref{Fig1}(b) shows the exclusivity graph of the 8 events in the Bell-CHSH inequality experiment performed in Stockholm, Fig.~\ref{Fig1}(c) shows the exclusivity graph of the 8 events in the NC inequality experiment performed in Rome and Fig.~\ref{Fig1}(a) shows a subgraph of the exclusivity graph of the set of $8 \times 8$ global events resulting of considering Stockholm and Rome's as two parts of a single experiment. The entire exclusivity graph of the two-city experiment is shown in Fig.~\ref{Fig4}.

%%%%%%%%%%%%%%%%%%%%%%%%%%%%%%%%%%%%%%%%%%%%%%%%%%%%%%%%%%%%%%%%%%%
% Fig. 4
%%%%%%%%%%%%%%%%%%%%%%%%%%%%%%%%%%%%%%%%%%%%%%%%%%%%%%%%%%%%%%%%%%%

%%%%%%%%%%%%%%%%%%%%%%%%%%%%%%%%%%%%%%%%%%%%%%%%%%%%%%%%%%%%%%%%%%%

\end{document}